# The lock and key model for Molecular Recognition.

# Is it time for a paradigm shift?

Arieh Ben-Naim

Department of Physical Chemistry

The Hebrew University of Jerusalem

Givat Ram, Jerusalem 91904

Israel

## Abstract

We review the standard lock and key (LK) model for binding small ligands to larger adsorbent molecule. We discuss three levels of the traditional LK model for binding. Within this model the binding constant or the Gibbs energy of the binding process is related to the total interaction energy between the ligand and the binding site of the adsorbent molecules. When solvent molecules are present, which is the case in all binding processes in biochemistry, we find that a major part of the Gibbs energy of binding could be due to interactions mediated through the solvent molecules. This finding could have major consequences to the applicability of the LK model in drug design, and perhaps require a shift in the prevailing paradigm in this field of research.

## 1. Introduction

Binding processes are ubiquitous in biological systems [Ritter (1996), Tropp (1997), Stryer (1975), Ben-Naim (2001, 2010). These range from binding small molecules like oxygen to hemoglobin, various drugs to protein or to DNA, to binding of proteins to DNA [Hard and

2Lundback (1996), Ptashne (1967), Helene and Lancelot (1982), von Hippel (1994), von Hippel and Goldberger (1979), von Hippel and Berg (1986)]. In all of these the binding mode is highly specific. In this article we discuss the binding of a small ligand to a large macromolecule. This can be extended to the more general process of self-assembly of macromolecules, which we will not discuss here.

The idea that binding phenomena are controlled by the co-called Lock and Key (LK) model is quite old. It is attributed to Emil Fischer who postulated this model in 1894. The idea is very simple; the specific action of an enzyme on a substrate can be explained using a Lock and Key analogy. In this analogy, the lock is the enzyme and the key is the substrate. The enzymatic activity can occur only when the correct form of the key (substrate) fits into the key-hole (active site) of the lock (enzyme). The main idea is very simple and easy to understand. A ligand (L) binding to an absorbent molecule (A) will bind more tightly when L "fits" better into the binding site on A.

Stillinger and Wasserman (1978), Tabushi and Mizutani (1987), Rebek et al (1987,1988), Lightner et al (1987). Ben-Naim (2001).

"Fitting," in its original sense meant *geometrical fitting*, much as the fitting of a key to a lock, Figures 1.

As with real keys and locks, the idea of fitting has gone through a series of variations. In old keys, the *shape* of the key had to fit the shape of the keyhole for it to be able to work1. Later, the fitting meant to be between a pattern along the key and the counter pattern within the lock. In today's technology, remote control keys do not have any patter fitting at all. The key sends an electronic signal which enables the locking, or the unlocking of the lock.

Likewise, the LK model for binding has gone through a series of modifications. From a simple geometrical fitting, Figure 2, to group-pattern fitting, induced fitting and finally, doing away with the fitting altogether.

In the next section (2), we shall review the various LK models, and their definitions in terms of the Gibbs energy of binding. In section (3), we shall discuss the new paradigm for binding which does not depend on the geometrical fitting between the ligand and the adsorbent molecule. We shall also discuss some of the consequences of the new paradigm for the theoretical aspects of drug design.

## 2. The various versions of the LK model

### 2.1. Purely geometrical fit

Figure 3 shows a hard sphere ligand L and a hard adsorbent molecule A having three different sites, 1, 2 and 3.

By hard particles we mean that two particles interact via a hard-pair-potential which is infinitely repulsive at distances smaller than some distance $\sigma$ and zero everywhere else, Figure 4. In this case, although the ligands "fit" better to site 2, its binding energy is the same for any other site on the surface of A. In other words, the *pure geometrical* fit does not imply that L will preferentially bind to site 2. The binding Gibbs energy of L to any of the site is zero.

### 2.2 Geometrical fit with weak interaction energy

In most real cases of ligands binding to sites the geometrical fit means maximum interaction energy between the ligand and the groups of the absolvent molecule at the site.

In this case the "recognition" of the binding site has been believed to be achieved through *direct* interaction between the ligand and the site. In this view the selection of the binding site is according to the criterion:

$$Min[\Delta U(i)] \qquad (2.1)$$

Where $\Delta U(i)$ is the direct interaction between the ligand and the *i*th site on the polymer A. The minimum is over all possible binding sites, $i$ on A.

The simplest means of recognition is through the weak van der Waals interactions. The criterion (2.1) is fulfilled whenever there are more groups on L that interact with groups on A. This is equivalent to the largest area of the surface of contact between L and A, hence the *geometrical* fit, which characterized the lock-and-key model.

Figure 5 shows three ligands $L_1, L_2$ and $L_3$, and an adsorbent molecule A having three geometrically different sites, $A_1, A_2$ and $A_3$. Without doing any calculation we can correctly guess that ligand $L_1$ will preferentially bind to site $A_1$, ligand $L_2$ will preferentially bind to site $A_2$, and ligand $L_3$ will preferentially bind to site $A_3$.

Underlying this guesswork is the assumption that in the absence of the solvent, and for all rigid molecules the *Gibbs energy* of binding will be reduced to the energy of binding, Ben-Naim (1992, 2001, 2010) i.e.

$$\Delta G(binding) = \Delta U(binding) \qquad (2.2)$$

Therefore, also the binding constants and the probability of binding to a specific site will be proportional to

$$\Pr(\text{binding to a specific site}) = C \exp[-\beta \Delta U(binding)] \qquad (2.3)$$

where C is a normalization constant and $\beta = (k_B T)^{-1}$, with $k_B$ the Boltzmann constant and T the absolute temperature.

Clearly, because of equation (2.3) we can identify the geometrical fit with the maximal number of interacting pairs of groups one belonging to the ligand and one belonging to the specific sites, see Figure 6. For instance, ligand $L_1$ would have about 10 group-group interactions with site $A_1$, about 4 interaction energies with site $A_2$ and about 2 interaction energies with site $A_3$. Assuming that each pair interaction contributes the same quantity $u$ to the binding energy, we can conclude that:

$$\Delta U(L_1 \text{ on } A_1) = 5u$$

$$\Delta U(L_1 \text{ on } A_2) = 3u$$

$$\Delta U(L_1 \text{ on } A_3) = 2u \qquad (2.4)$$

Since $u$ is a negative quantity we can conclude that:

$$\Pr(L_1 \text{ on } A_1) > \Pr(L_1 \text{ on } A_2) > \Pr(L_1 \text{ on } A_3) \qquad (2.5)$$

which means that $L_1$ will preferentially bind to $A_1$, less to $A_2$, and less to $A_3$, Figure 6.

Similarly, for $L_2$ we can calculate the approximate number of contacts between groups on $L_2$ and groups on the sites of A and conclude that, see Figure 7.

$$\Pr(L_2 \text{ on } A_2) > \Pr(L_2 \text{ on } A_1) > \Pr(L_2 \text{ on } A_3) \qquad (2.6)$$

Similarly for $L_3$ we can calculate that it will preferentially bind to $A_3$.

This is essentially the physical reason underlying the LK model for binding; geometrical fit means stronger interaction energy, hence also higher probability to bind to a specific site,

It should be emphasized that by interaction energies we mean here *direct* interaction energy between groups on the ligand and groups at the specific site on A. Before we describe the solvent effect of binding we mention two more variations of the LK model which also depend on *direct* interactions.

## 2.3 Fitting by means of complimentary functional groups

Figure 8 shows a ligand L having three functional groups; a hydroxyl group (OH), and a non-polar group (say methyl) and a positive charged group. The adsorbent molecule has three possible sites each having three functional groups. Clearly, because (+) is attracted to (-), and OH can form a HB with the CO, the ligand will bind preferentially to site $A_1$ rather to either site $A_2$ or $A_3$. The reason is that the charge-charge, and the hydrogen bonding on site A involves the strongest *direct* interaction energy between L and the site. Hence, we have:

$$\Pr(L \text{ on } A_1) > \Pr(L \text{ on } A_2) > \Pr(L \text{ on } A_3) \qquad (2.7)$$

Note however, that geometrical fitting does not play any role here. Only the *direct* interaction energies between the various groups on the ligand, and on the site determine the interaction energy, hence the preferential probabilities.

Another variation of the lock-and-key model is the so-called induced-fit-model, which is essentially the same as the ones described above, but here the fit is achieved after the binding has occurred, Figure 9.

The criterion (2.1) may be translated into probabilities, provided the process of binding is carried out in the absence of a solvent. For instance, having two sites (a) and (b), we can say that the probability of binding to the site (a) is larger than the probability of binding to site (b). The relationship between the probability ratio and the difference in the binding energies is

$$\frac{\Pr(a)}{\Pr(b)} = \frac{\exp[-\beta \Delta U(a)]}{\exp[-\beta \Delta U(b)]} = \exp[-\beta \{\Delta U(a) - \Delta U(b)\}] \qquad (2.8)$$

This ratio is also equal to the ratio of the two measurable binding constants. Note that the two carbonyl groups are far apart and therefore the interaction between them is negligible in the absence of a solvent.

## 3. Molecular recognition through the solvent

When the binding occurs in a solvent, the criterion for the selection of the preferable binding site is not (2.1) but instead:

$$Min[\Delta G(i)] \tag{3.1}$$

Where $\Delta G(i)$ is the Gibbs energy change for binding to the *i*th site. The difference between $\Delta G(i)$ and $\Delta U(i)$ is the solvent-induced contribution to the binding Gibbs energy, and this is related to the solvation Gibbs energies of the three molecules L, A and AL, i.e.

$$\delta G(i) = \Delta G_{AL}^* - \Delta G_A^* + \Delta G_L^* \tag{3.2}$$

In simple non-aqueous solvents the main contributions to $\delta G(i)$ comes from the Hard (H) and the Soft (S) parts of the solute-solvent interactions. As we have seen in section 2 these two contributions depend on the volumes and on the surface areas of the three solutes involved in the binding process. It is commonly believed that even when $\delta G(i)$ is large compared with $\Delta U(i)$, the difference in the values of

$\delta G(i)$ for different sites *i*, might not be very large. In other words, the modified criterion (3.1) might be very different from the criterion (2.1), but the *difference* between the two might not be sensitive to the specific site. Ben-Naim (1980, 1987a, 1987b, 1990,1992, 1994). As an extreme example suppose that for each site *i*, we have

$$\Delta G(i) = \Delta U(i) + \delta G \tag{3.3}$$

Where $\delta G$ is independent of *i*. In such a case, no matter how large $\delta G$ might be, its effect on the probability ratio for binding to any two sites (a) and (b) would be negligible, i.e.

$$\frac{\Pr(a)}{\Pr(b)} = \frac{\exp[-\beta \Delta G(a)]}{\exp[-\beta \Delta G(b)]} \approx \frac{\exp[-\beta \Delta U(a)]}{\exp[-\beta \Delta U(b)]} \tag{3.4}$$

As we noted earlier the probability ratio is equal to the ratio of the experimentally measurable binding constants to the sites (a) and (b).

Thus, although the binding constant might be considerably modified in the presence of the solvent, the relative preference for the binding site might be the same as in the gaseous phase. This is equivalent to the statement that the lock-and-key model is still valid; the binding Gibbs energy is modified, but the preferential binding site is still determined by the total direct interaction energies $\Delta U(i)$.

The argument given above, whether made explicitly or implicitly still dominates the thinking of scientists in the pharmaceutical sciences who are engaged in drug design. However, recently it was pointed out that in some binding processes occurring in aqueous solutions, the involvement of hydrophilic ($H\phi I$) effects might be so profound that the lock-and-key model might be rendered completely irrelevant to the selection of preferential site. In some cases it can thoroughly modify the way one approaches the problem of drug design, Ben-Naim (1987, 2002, 2006, 2009, 2011), Wang and Ben-Naim (1996).

We present here one example where solvent-induced effect, based on $H\phi I$ interaction can *reverse* the preference for the binding sites.

Consider again the two sites (a) and (b) in Figure 10. Clearly, the ligand L fits better to site (a) than to site (b). Thus, in accordance with the lock-and-key model site (a) will be preferred over site (b); in probability terms, the ration of the two probabilities is:

$$r^{(g)} = \frac{\Pr(a)}{\Pr(b)} = \frac{\exp[-\beta\Delta U(a)]}{\exp[-\beta\Delta U(b)]} > 1 \qquad (3.5)$$

Clearly, in the gaseous phase the ligand will prefer binding to site *a*. Equivalently, because of the stronger binding energies, the binding constant to site (a) will be larger than the binding constant to site (b).

The preferential binding site might be reversed in aqueous solution. Suppose that the ligand L and the polymer A each has a functional group (FG) that can form Hydrogen Bond (HB) with water. Note that these functional groups are not in the binding interface between L and A. Therefore, even in the presence of these FGs, the preferred binding site in the gaseous phase is still the site (a). This preferred site will probably be maintained if we add an organic solvent, say hexane or benzene. However, in aqueous solutions the whole story is quite different. Suppose that the FGs on L and A are such that when binding to (b), they can be bridged by a water molecule. In this case, the binding Gibbs energy to site (b) will be modified:

$$\Delta G(b) = \Delta U(b) + \delta G(b) \qquad (3.6)$$

In binding to site (a) the two FGs are far apart so that they do not interact directly or indirectly, therefore the binding Gibbs energy to (a) is:

$$\Delta G(a) = \Delta U(a) + \delta G(a) \qquad (3.7)$$

where we have:

$$|\delta G(a)| << |\delta G(b)| \qquad (3.8)$$

The probability ratio in the aqueous solution is:

$$r^{(w)} = \frac{\Pr(a)}{\Pr(b)} = \frac{\exp[-\beta \Delta G(a)]}{\exp[-\beta \Delta G(b)]} = r^{(g)} \frac{\exp[-\beta \delta G(a)]}{\exp[-\beta \delta G(b)]} \qquad (3.9)$$

For simplicity assuming that $\delta G(a)$ is negligible, and that $\delta G(b)$ is due to one $H\phi I$ correlation which amount to about $-5\, k_B T$. In this case we shall have

$$r^{(w)} \approx \frac{r^{(g)}}{150} \qquad (3.10)$$

Thus, although we have started with $r^{(g)} > 1$, i.e. the preferential site is (a), the addition of one $H\phi I$ interaction could change the probability ratio by a factor of 150 in favor of the site (b). Clearly, in such a case, the whole lock-and-key argument becomes irrelevant to the problem of preferential binding site.

The finding that $H\phi I$ interactions can change the preferential binding site has far reaching consequences to the problem of drug design, either for designing new drugs or for modifying existing drugs to improve their efficacy. (See for example: Kuntz (1992), Perun and Propst (1989), Propst and Perun (1992), Greer et al (1994) and Roerding and Kroon (1989))

Some specific examples were discussed recently. We shall not present these highly technical examples here. The interested reader should consult the article by Wang and Ben-Naim (1996).

In connection with the solvent-induced effect on preferential binding, it should be noted that proteins' surface are very in-homogenous, both in their structure and their chemical constituency. Therefore, it is likely that in most binding processes to protein, both the direct (i.e. lock-and-key model), and the indirect interactions are of comparable magnitudes. There might be crevices on the surface of the proteins that provide tightly fitted binding sites, as well as $H\phi I$ interaction through $H\phi I$ groups that do not belong to the binding interface. The situation is quite different in DNA, where the surface is much more homogenous, both in terms of the structure as well as in terms of the distribution of FGs. Therefore, in the binding of ligands, drugs or even proteins to DNA, it is more likely that the $H\phi I$ interaction will dominate the binding Gibbs energy, and hence also the selection of the binding site.

We next turn to a hypothetical example, where we have a seemingly featureless surface, for instance a flat surface with $H\phi I$ groups randomly distributed on it. Looking superficially on this surface, we might not detect any preferred binding site to the ligand L, Figure 11a. Therefore, from the point of view of the lock-and-key model, there exists no preferential binding site.

A ligand approaching this surface in water, might "feel" very different affinities to different regions on this apparently homogenous surface. A ligand hovering above the surface might find one region far more favorable for binding than any other region. An illustration of such an example is shown in Figure 11b.

Let Pr(0) be the probability of binding of the ligand to an arbitrary region on the surface of the protein which does not contain $H\phi I$ groups Examples (1), (2) and (3) in Figure 11. In the absence of a solvente, the Gibbs energy of binding to any point of these regions on the surface of the protein is nearly independent of the location of the binding site. However, in water the ligand might find sites at which the binding Gibbs energy will be much larger due to the formation of one, two or three $H\phi I$ interactions, by means of a water-bridge. The relative probabilities for binding to sites (1), (2) and (3), Figure 11 are:

$$\frac{\Pr(1)}{\Pr(0)} \approx \exp[5] \approx 150$$

$$\frac{\Pr(2)}{\Pr(0)} \approx \exp[10] \approx 2.2 \times 10^4$$

$$\frac{\Pr(3)}{\Pr(0)} \approx \exp[15] \approx 3.3 \times 10^6 \tag{3.11}$$

It is clear from this example that a seemingly featureless surface might provide binding sites, such as (1), (2) and (3) with significantly different binding constants. We can also imagine that a ligand approaching a DNA molecule might "see" a nearly homogenous surface on the DNA. However, looking through the water, the ligand might "see" some sites which are more preferred for binding than the others.

We have discussed here only one type of solvent induced interactions. There are many other possibilities; some of longer range and some of stronger interactions. Longer range can be achieve by chain of two or more molecules figure 12, stronger interactions can be achieved by water molecule bridging more than two functional groups Ben-Naim (2011)

## 4. Conclusion

For over a hundred years the lock and key model for binding a ligand to a site was the only model for binding both small ligands (such as substrate to an enzyme) and large proteins (binding to DNA). Even when a solvent was present, the lock and key model was considered to be valid. The theory of binding did not change. Perhaps the solvent would modify the strength of the interaction between the ligand and the site, hence the binding constant would be affected, but the theory itself was unaffected.

Having discovered the importance of the hydrophilic interactions, i.e. water forming hydrogen-bond bridge between two or more hydrophilic groups, the argument based on the lock and key mechanism will not, in some cases be relevant to the binding mechanism. For these cases we suggest that it is time to a paradigm shift from the lock and key model, to specific solvent-induced effect on the binding of a ligand to

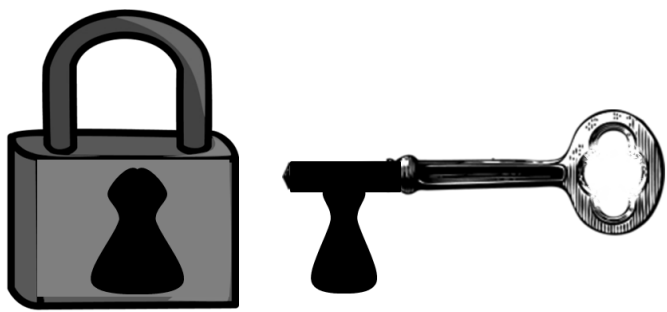

Figure 1.   An old key fitting geometrically to the entrance of the lock

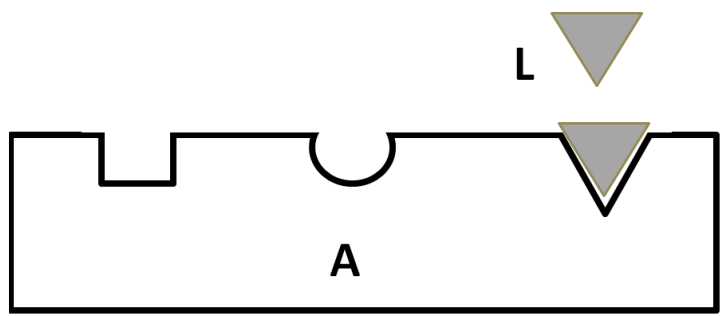

Figure 2.  The geometrical fit between a key L (substrate) and a specific site on the lock A (enzyme)

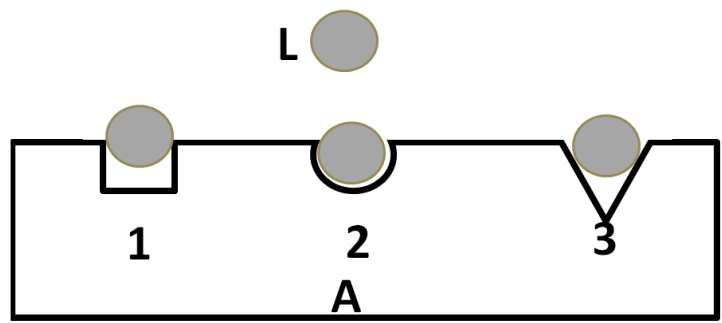

Figure 3. A hard sphere ligand L perfectly fits the site 2 on the lock A. In this case better fitting does not imply better binding

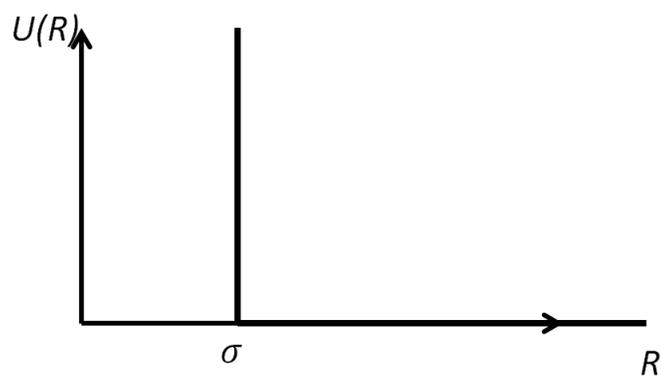

Figure 4. The interaction energy, $U(R)$ between two hard spheres of diameter $\sigma$ as a function of the distance $R$.

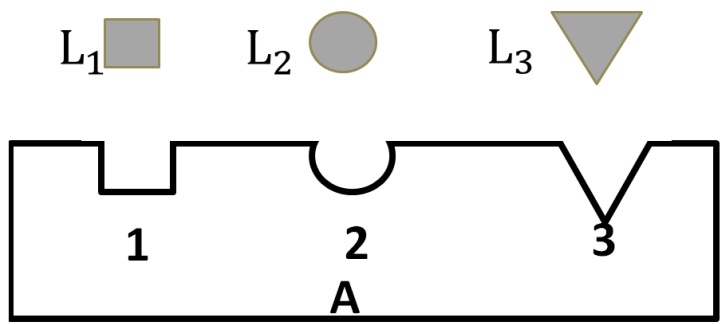

Figure 5. The interaction energy between two hard spheres of diameter $\sigma$ as a function of the distance $R$

 

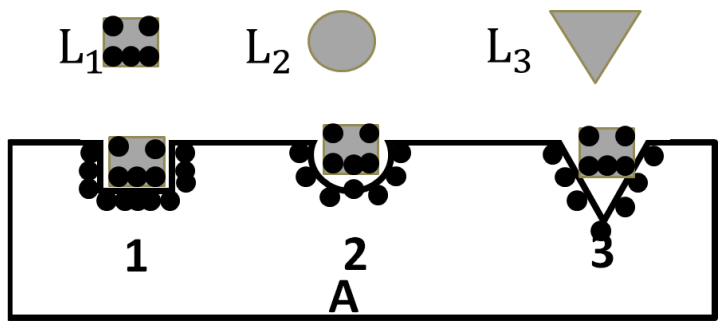

Figure 6. The interaction energy between ligand $L_1$ and site 1 will be stronger that to either site 2 or 3

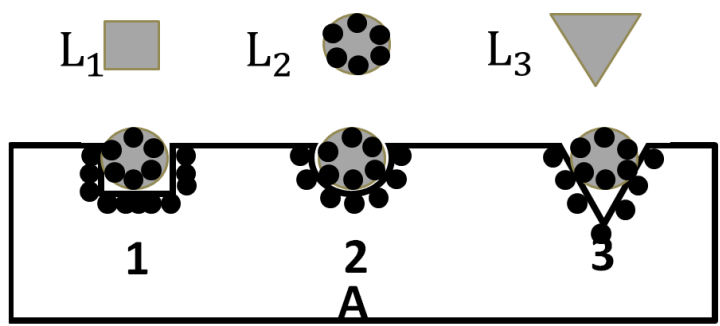

Figure 7. The interaction energy between ligand $L_2$ and site 2 will be stronger that to either site 1 or 3

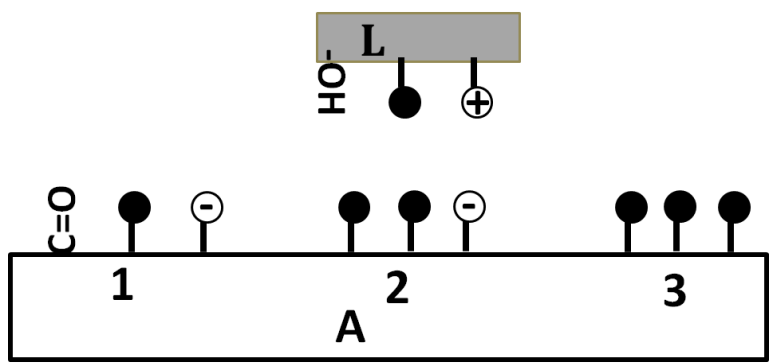

Figure 8. The interaction energy between ligand L and site 1 will be stronger that to either site 2 or 3

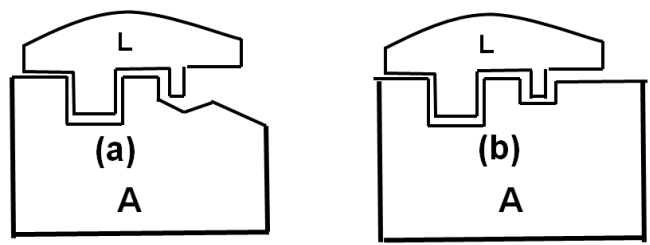

Figure 9. (a) L Binds to A, (b) after the binding the conformation of A changes to better fit the structure of L

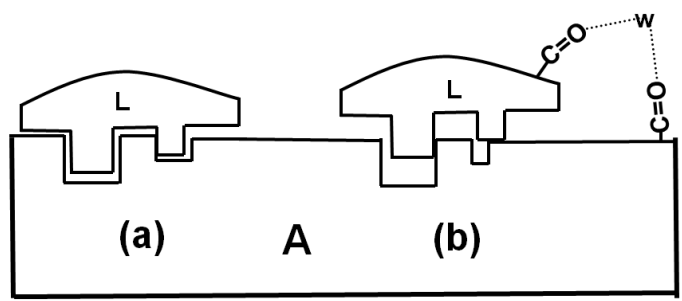

Figure 10. The ligand L fits better to site a than to site b. In the presence of a solvent (water), the ligand might prefer site b over a

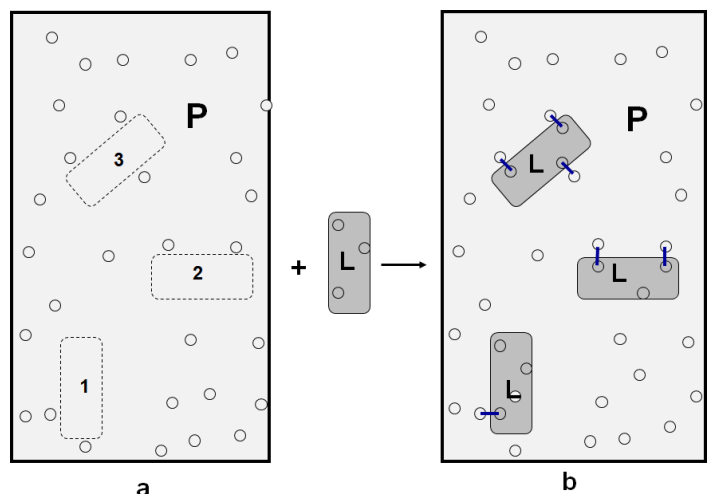

Figure 11. An apparent featureless surface (a), may offer preferred binding sites (b)

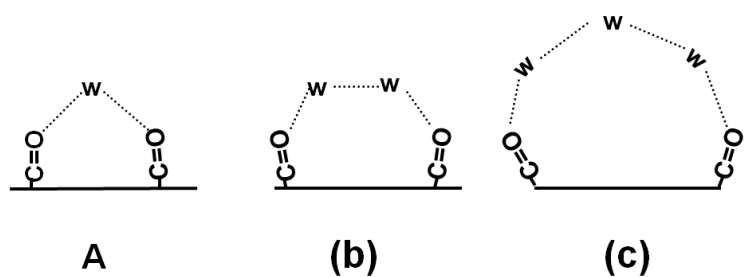

Figure 12. Longer range of solvent induced interaction, by means of (a) one-water bridge, (b) two-water bridge and (c) three-water bridge